# A coupled CFD-FEA study of the sound generated in a stenosed artery and transmitted through tissue layers

Fardin Khalili*, Peshala P. T. Gamage, Ibrahim A. Meguid, and Hansen A. Mansy, *Member*, IEEE
Biomedical Acoustics Research Laboratory, University of Central Florida, Orlando, FL 32816, USA
{fardin@knights., peshala@knights., hansen.mansy@} ucf.edu

*Abstract*—A new computational approach for simulating the blood flow-induced sound generation and propagation in a stenosed artery with one-sided constriction was investigated. This computational hemoacoustic method is based on mapping the transient pressure (force) fluctuations on the vessel wall and solving for the structural vibrations in frequency domain. These vibrations were detected as sound on the epidermal surface. The current hydro-vibroacoustic method employs a two-step, one-way coupled approach for the sound generation in the flow domain and its propagation through the tissue layers. The results were validated by comparing with previous analytical and computational solutions. It was found that the bruits (generated from the flow around the stenosis) are related primarily to the time-derivative of the integrated pressure force on the arterial wall downstream of the stenosis. Advantages of the methods used in the current study include: (a) capability of providing accurate solution with a faster solution time; (b) accurately capturing the break frequency of the velocity fluctuation measured on epidermal surface; (c) inclusion of the fluid–structure interaction between blood flow and the arterial wall.

*Keywords—Heart sound, cardiovascular flow, hemodynamics, hemoacoustics, bruits, murmurs.*

## I. Introduction

Cardiac Echocardiography employs Doppler ultrasound for the assessment of intracardiac flows, and Phonocardiography employs recording and analysis of heart sounds at the skin surface [1–7]. Manual auscultation has been used for many decades for cardiovascular disease diagnostics [8,9]. Blood flows associated with many abnormal cardiovascular conditions generate characteristic sounds called "murmurs" or "bruits" [10]. These sounds can be measured on the skin surface using a stethoscope [11]. However, the physical mechanisms that generate these sounds, as well as the physics of sound transmission through the body, are still not well understood [12]. It has long been accepted that the source of most murmurs are disturbances in blood flow caused by obstruction in the vessels. Modeling of these structures would be helpful when more detailed flow behavior such as in studies of acoustic sources is needed [13]. Hence, there have been many previous studies on the dynamics of flows through stenosed or partially obstructed vessels [14,15]. In addition, there have been a few modeling studies employing finite-element [16] or boundary-element based methods [17] on wave propagation in tissue-like materials, but these studies were conducted for highly simplified cases with prescribed sources. In order to more fully understand the relationship between cause (disease) and effect (sound measured on the skin surface), the hemodynamics associated with the murmur must be investigated concurrently, while considering the complete elastic wave dynamics including compression and shear waves propagation, and wave scattering and dissipation. The direct simulation of blood flow-induced sounds has the potential to provide an unprecedented understanding of heart murmurs, and this forms the primary motivation for the present study. Similar investigations has been done in computational aeroacoustics (CAA) and hydroacoustics fields [18–20].

Several studies were performed to understand the source and mechanism of the bruit generation. Bruns [10] argued that arterial bruits were generated by the 'nearly periodic fluctuation in the wake found downstream of any appropriate obstacle' and not by post-stenotic turbulence. Lees and Dewey [21] recorded the spectrum of actual bruit sounds (a technique called phonoangiography), and suggested a significant similarity between the bruit sound spectrum and the wall pressure spectrum of a fully developed turbulent pipe flow. Duncan [22] developed a relationship to estimate the residual lumen diameter of a stenosed area based on the break frequency observed in the measured bruit spectra. Fredberg [23] derived a theoretical model for the transfer function between wall pressure spectrum and sensed sound using the Green's function and a stochastic analysis of turbulent boundary layer. Wang et al. [24] modeled the sound generation in a stenosed coronary artery using an electrical network analog model, and Borisyuk [25] modeled the sound propagation through the tissues (thorax) theoretically for a simple cylindrical geometry. In a recent study, the blood flow-induced arterial ''bruits'' were computed directly using a hybrid approach wherein the hemodynamic flow field is solved by an immersed boundary, incompressible flow solver, and the sound generation is modeled based on the linearized compressible perturbation equations [12,26]. The transmission and propagation of the sound through the surrounding biological tissues is also modeled with a simplified, linear structural wave equation.

Computational modeling offers a promising modality for exploring the physics of heart murmurs. Thus, virtual Echocardiography (ECHO) and Phonocardiography (PCG) can

*Corresponding author.
This study was supported by NIH R44HL099053.

XXX-X-XXXX-XXXX-X/XX/$XX.XX ©20XX IEEE

serve as a bridge between clinical data and computational hemodynamic results [27,28]. These virtual ECHO and PCG can be used for rapid validation of computational results by comparing to the actual cardiographic data, and furthermore, such comparisons will allow us to calibrate a given patient-specific, computational heart model. In the current paper, we present a computational method for the direct simulation of the generation of blood flow-induced sounds and the propagation of these waves through a tissue-like material. Blood flow in the vessel was simulated by solving the incompressible Navier–Stokes equations and propagation through the surrounding vessel wall and tissue is resolved with a "harmonic response" finite element analysis (FEA). The pressure fluctuations causing vibrations in the solid domain were investigated to resolve wave propagation and scattering accurately. The flow field inside the artery and the bruit sound signal at the epidermal surface were examined to delineate the source of the arterial bruit and the correlation between the bruit and the arterial wall pressure fluctuations.

## II. METHODS

Most fluid flows are characterized by irregularly fluctuating flow quantities that often occur at small scales and high frequencies. Hence, resolving these fluctuation in time and space requires excessive computational cost. Optimum modeling of these structures [13,29,30] is of interest for the acoustic investigations including biomedical applications, which are active areas of research [2–4,31–34].

### A. Models

A two-dimensional constricted channel is considered (Fig. 1) and is the same as in a previous study [26]. This geometry serves as a model of a stenosed artery in patients with vascular diseases.

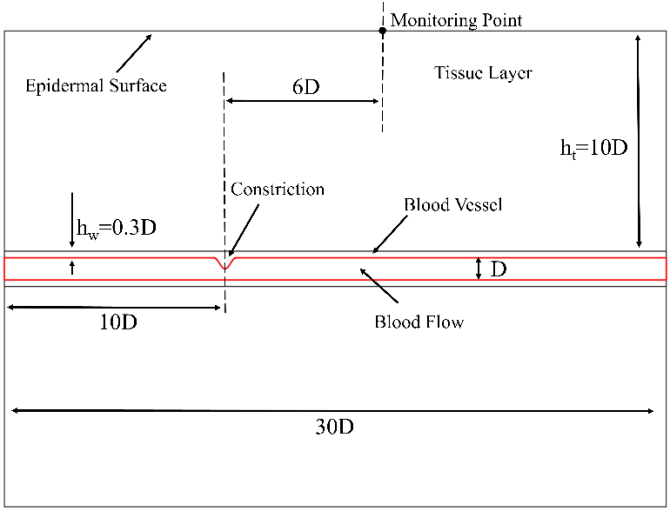

Fig. 1. Schematic of the constricted channel model and acoustic domain; D arterial diameter, $h_w$ arterial wall thickness; $h_t$ tissue layer thickness.

The channel is constricted from top wall, and the profile of the constriction is given by

$$y = y_{max} - \frac{b}{2}\left[1 + cos\left(2\pi \frac{x - x_0}{D}\right)\right]; \quad (1)$$

where, $x_0$ is the center of the stenosis, $D = 5$ mm is the height of the channel, and $b = 0.5D$ is the size of the constriction. Also, $-D \leq (x - x_0) \leq D$. Similar constricted artery models were used in earlier studies [35,36].

### B. Hemodynamics

The CFD analysis was performed for a pulsatile flow. The flow was driven by a pulsatile pressure drop between the inlet and exit (shown in Fig. 1) with the following sinusoidal variation in time (non-dimensional form):

$$\frac{\Delta P}{\rho U^2} = [A + B \sin(2\pi ft)] \quad (2)$$

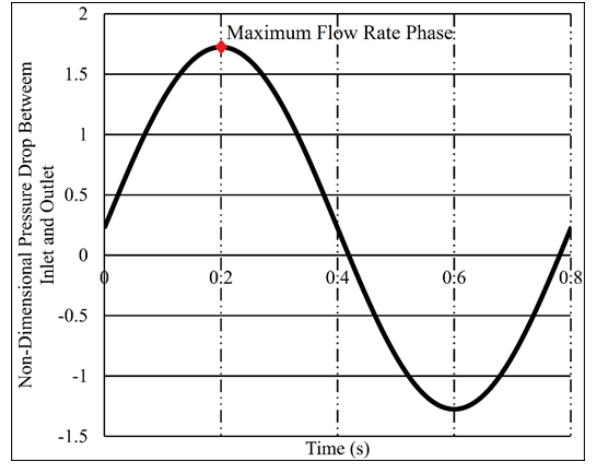

Fig. 2. Non-dimensional pulsatile pressure drop between inlet and outlet.

where, A is set to 0.225 and B to 1.5. The non-dimensional frequency of pulsation is $St = fD/U_{max} = 0.048$ and the Reynolds number is set to $Re = \rho U_{max} D/\mu = 1000$, where $f = 5$ Hz is the frequency of pulsatile flow, $U_{max}$ is the maximum centerline velocity at the inlet, density $\rho = 1050$ kg.m$^{-3}$, and dynamic viscosity $\mu = 0.0035$ Pa.s. The chosen flow parameters yield a Womersley number which is in the appropriate range for large arteries [37] equals to $\alpha = (\pi \cdot Re \cdot \frac{St}{2})^{1/2} = 8.6$. The Wilcox's standard-Reynolds k-ω turbulence model [38–42] was used to simulate the flow, which is known to perform well for internal flows [43–48]. The unsteady simulation was performed with a time step of 0.1 ms and 10 iterations per time step. Numerical solution typically converged to residuals about $< 10^{-3}$. Moreover, high quality triangular mesh was generated in the flow domain, especially in the stenosed region. Therefore, y+ was maintained less than 1 close to all walls. In the current model, the blood flow is assumed to be Newtonian (which is a good assumption for the larger and medium sized arteries [32]). Hence, the governing equations are the following incompressible Navier–Stokes equation,

$$\rho \frac{\partial u_i}{\partial t} + \rho u_j \frac{\partial u_i}{\partial x_j} = -\frac{\partial p}{\partial x_i} + \mu \frac{\partial^2 u_i}{\partial x_j \partial x_j} \quad (3)$$

where, $u_i$ is velocity vector, p is pressure, and $\rho$ is the blood density. In addition, a mesh-independent study was conducted to find the optimized mesh configuration. Prism layer mesh was also employed near the boundaries since accurate prediction of pressure drop in flows with separation depends on resolving the velocity gradients normal to the wall [49,50]. Dirichlet pressure boundary conditions are applied at the exit ($p_{exit} = 0$), and a no-slip boundary condition is used for the top and bottom walls. The flow computations are carried out for about four pulsation cycles after it reaches a stationary state.

*C. Acoustic*

The transient pressure (force) fluctuations on the vessel wall excite the solid domain, causing vibrations which in turn detected as sound on the epidermal surface. Harmonic Response Analysis (HRA) in ANSYS finite element analysis (FEA) software package was used to simulate these vibrations. HRA calculates the steady-state (harmonic) response of a linear structure subjected to a harmonically varying load. HRA can solve for the response of a structure to harmonically varying loads over a frequency range. The equation of motion of a structure under harmonic loading can be derived as,

$$M\ddot{x} + C\dot{x} + kx = F(t) \quad (4)$$

$$F(t) = F_{max}e^{i(\omega t + \varphi)} \quad (5)$$

$$x(t) = x_{max}e^{i(\omega t + \theta)} \quad (6)$$

$$(-M\omega^2 + i\omega C + k)x_{max}e^{i(\omega t + \theta)} = F_{max}e^{i(\omega t + \varphi)} \quad (7)$$

In above equations, Eqn. (4) represents the equation of motion of a structure in time domain, where $M$, $C$, and $k$ denote structural mass, damping and stiffness matrices. If the applied force, $F(t)$, is harmonic, it can be represented as in Eqn. (5) where, $F_{max}, \omega, \varphi$ are the force amplitude, angular frequency and phase shift, respectively. Similarly, the displacement $x(t)$ is also harmonic under harmonic loading and presented in Eqn. (6), where $x_{max}$, $\theta$ are the magnitude and phase shift of displacement, respectively. By substituting Eqn. (5) and (6) in Eqn. (4), the derived Eqn. (7) is solved in HRA simulation. In the present work, the pressure fluctuations on the vessel wall are recorded from the CFD solution, which are later transformed in to frequency domain using Fast Fourier Transformation (FFT). The transformed pressure in the frequency domain are then mapped on to the vessel wall in HRA simulation.

Fig. 3 shows the simulation methodology in the current study:

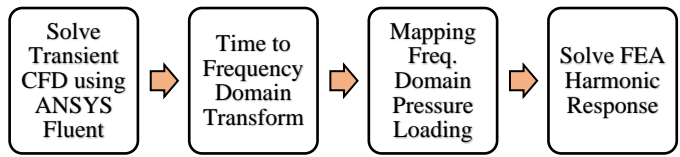

Fig. 3. Simulation methodology of the hydro-vibroacoustic method using ANSYS Fluent and FEA Harmonic Response.

In the current study, the fluid–structure interaction of blood flow with the arterial wall was also considered; however, it was neglected in the previous study [26]. These interactions with the elastic blood vessel may introduce resonance peaks in the sound spectrum [51]. However, these resonance peaks are generally vanished due to the damping associated with the surrounding tissue and may not have important components in the sounds detected at the skin surface [26,51]. The acoustic domain in the current study includes not only the lumen surface but also the arterial wall (blood vessel) and the surrounding tissue layers. The acoustic material properties were: the density of 1050 kg/m$^3$, 1100 kg/m$^3$, 1200 kg/m$^3$, and speed of sound of 1500 m/s, and 1580 m/s, and 1720 m/s for the blood, vessel wall and surrounding tissue, respectively. The top boundary of the acoustic domain represents the epidermal surface at which a stethoscope can sense transmitted sound via the displacement, velocity, or acceleration of the epidermis. It is also assumed that the acoustic waves radiate through all other boundaries. The shear waves generated in the tissue were considered negligible compared to the acoustic waves and that the viscous dissipation of the acoustic wave was also neglected [18,26,28]. In a previous study [26], only the bulk modulus and speed of sound of the materials were specified. However, the same bulk modulus corresponds to many combinations of Young's modulus and poison's value. The latter parameters highly affect the stiffness of the vessel wall and tissue layers, which consequently alter the amplitude of the sound propagation. Hence, each bulk modulus can correspond to many solutions. In addition, any difference in the Reynolds number could highly affect the flow behavior and the amplitude of the pressure forces on the vessel wall.

### III. RESULTS & DISCUSSION

The instantaneous vorticity contours at different times during one cycle are shown in Fig. 3. The results are in good agreement with the previous results [26,37].

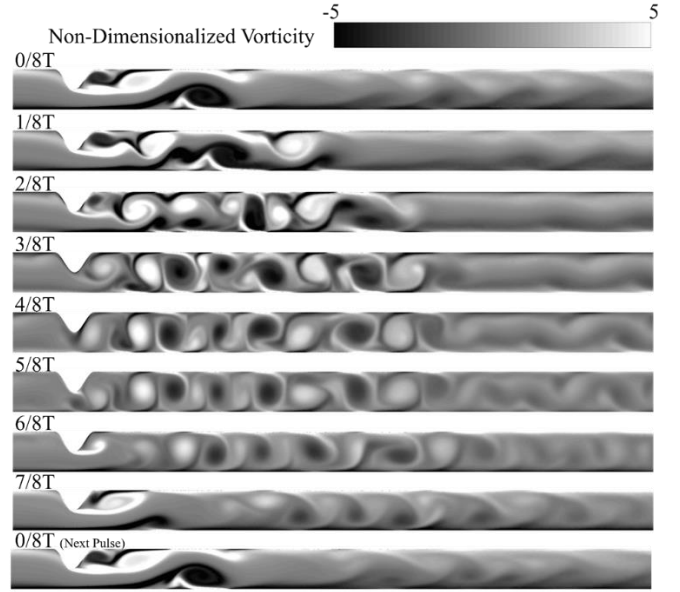

Fig. 3. Time evolution of vorticity field; 0/8T maximum flow rate, 4/8T minimum flow rate phase. (The vorticity contours shown were normalized by time scale D/U$_{max}$).

For the 50 % stenosed artery, it is observed that the vortex roll-up starts from the maximum flow rate phase (0/8T, where T is the period of pulsation). The detachment of separation bubble in the wake of the stenosis, and the boundary layer separation at the bottom surface are clearly visible. At 4/8T, the shear layers become unstable during deceleration and a coherent vortex series are formed as shown with an overall wavelength of about 1D. In this figure, 0/8T and 4/8 corresponds to maximum and minimum flow rates, respectively. In addition, the vorticity was normalized by time scale $D/U_{max}$, and scaled from -5 to 5.

The signal through the tissue layer was monitored at 6D downstream from the stenosis on epidermal surface (see Fig 1), where the maximum acoustic energy was Observed. The frequency spectra and time-frequency of vertical velocity fluctuations ($v'$) are shown in Fig. 4. The vertical dashed lines in Fig. 4 indicates the break-frequency [22] where the slope of the spectrum changes significantly. The current model could accurately capture the break frequency of the spectrum at about 40 Hz. In addition to the break frequency, frequencies of ~1.25 Hz (the main frequency of the flow pulse) and 20 Hz similar to previous study were observed. The difference in the amplitude of the solid velocity and velocity fluctuations could be caused by the material properties of the vessel wall and tissue layers.

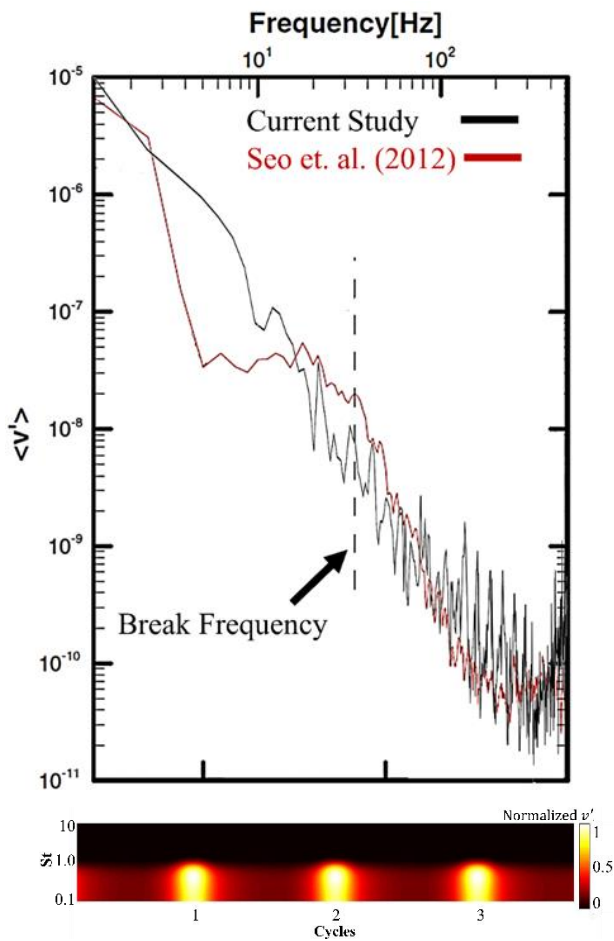

Fig. 4. Frequency spectrum of vertical velocity fluctuations ($v'$) on the epidermal surface monitored 6D downstream from the center of the stenosis. The non-dimensional frequency was shown on y-axis. The amplitude of the velocity fluctuations $v'$ of the tissue layer was also normalized.

The calculated spectra of vertical velocity fluctuations on epidermal surface using HRA agreed with the spectra calculated from linearized perturbation compressible equation [26]. In addition. The time–frequency distribution of the epidermal vertical velocity fluctuations was estimated using polynomial chirplet transform [52], which shows the intensity and frequency content of the arterial bruit.

## IV. Conclusions

A new computational approach for simulating the blood flow-induced sound generation and propagation in a stenosed artery with one-sided constriction was investigated. The advantages of the employed hydro-vibroacoustic method can be noted as:

- Considering both shear and longitudinal wave propagations.
- Accurately capturing the break frequency of the velocity fluctuations measured on epidermal surface.
- Capability of providing accurate solution with a faster solution time.
- Considering the fluid–structure interaction between blood flow and the arterial wall.

The analysis of the computed results indicated that the epidermal velocity fluctuations were correlated with transient pressure (force) fluctuations on the vessel wall more intense over the near post-stenotic region. This supports the view that the primary source of arterial bruits is the vortex induced perturbations in the near post-stenotic region.